\documentstyle[aps]{revtex}

\begin{document}
\author{Fayyazuddin}
\address{National Center for Physics, Quaid-i-Azam University,\\
Islamabad 45320, Pakistan}
\title{Some comments on $B\rightarrow \pi \pi $ decays}
\date{June 2004, NCP-QAU/2004-006}
\maketitle

\begin{abstract}
Isospin analysis has been used to constrain the CP-asymmetries in $%
B\rightarrow \pi \pi $ decays. In particular correlation between a weak
phase $\theta $ and a strong phase $\delta $ is obtained. Further using the
experimental values for the CP-averaged branching ratios, the following
bounds on direct CP-asymmetries are obtained: $-0.35\pm 0.22\leq C_{\pi
^{+}\pi ^{-}}\leq 0$; $C_{\pi ^{0}\pi ^{0}}=-\left( 2.4\pm 1.0\right) C_{\pi
^{+}\pi ^{-}}$. Constraints on mixing induced CP-asymmetries are also
discussed.
\end{abstract}

Isospin analysis has been used by several authors \cite
{PRL65-3381,PRD58-017504,PRD59-054007,PLB514-315} to constrain
CP-asymmetries in $B\rightarrow \pi \pi $ decays. There are five parameters
viz two amplitudes $f_{0}$ and $f_{2}$, two weak phases $\left( \theta
_{0},\theta _{2}\right) $ a strong phase $\delta _{0}-\delta _{2}$ to
describe all the observables of these decays. The observables for these
decays are three CP-averaged branching ratios $R_{+-}$, $R_{00}$, $R_{-0}$,
two direct CP-asymmetries $C_{\pi ^{+}\pi ^{-}}$, $C_{\pi ^{0}\pi ^{0}}$ and
two mixing induced asymmetries $S_{\pi ^{+}\pi ^{-}}$, $S_{\pi ^{0}\pi ^{0}}$
i.e. seven in numbers. Thus one would expect some constraints and
relationships among them. The purpose of this note to discuss these
constraints in the light of new data by BaBar and Belle collaborations \cite
{PRL89-201802,PRD66-071102,Jawahery2003}. Some overlap with the previous
work is expected; nevertheless there may be some new insight in the light of
new data. In a recent papers \cite{0312259,0401188} the $B\rightarrow \pi
\pi $ decays are also discussed.

As is well known, the decay amplitudes can be written in terms of two
complex amplitudes $A_{0}$ and $A_{2}$: 
\begin{eqnarray}
\bar{A}_{+-} &\equiv &A\left( \bar{B}^{0}\rightarrow \pi ^{+}\pi ^{-}\right) 
\nonumber \\
&=&\frac{1}{\sqrt{3}}\left[ \sqrt{2}A_{0}-A_{2}\right] =\frac{1}{\sqrt{3}}%
\left[ \sqrt{2}a_{0}e^{i\delta _{0}}-a_{2}e^{i\delta _{2}}\right] 
\label{e01} \\
\bar{A}_{00} &\equiv &A\left( \bar{B}^{0}\rightarrow \pi ^{0}\pi ^{0}\right) 
\nonumber \\
&=&\frac{1}{\sqrt{3}}\left[ A_{0}+\sqrt{2}A_{2}\right] =\frac{1}{\sqrt{3}}%
\left[ a_{0}e^{i\delta _{0}}+\sqrt{2}a_{2}e^{i\delta _{2}}\right] 
\label{e02} \\
\bar{A}_{-0} &\equiv &A\left( B^{-}\rightarrow \pi ^{-}\pi ^{0}\right)  
\nonumber \\
&=&-\sqrt{\frac{3}{2}}A_{2}=-\sqrt{\frac{3}{2}}a_{2}e^{i\delta _{2}}
\label{e03}
\end{eqnarray}
The following comments are in order: The two pions in the final state are in 
$I=0$, $I=1$ or $I=2$ isospin states. There are only two symmetric isospin
wave functions, one corresponding to $I=0$, with the final state phase $%
\delta _{0}$, the other corresponding to $I=2$ with the final state phase $%
\delta _{2}$. There are two antisymmetric wave functions corresponding to $%
I=1$, thus, there are two $I=1$ amplitudes with final state phases $\delta
_{1}^{\prime }$ and $\delta _{1}^{\prime \prime }$. The Bose statistics
excludes the antisymmetric wave functions and we are left with two
amplitudes $A_{0}$ and $A_{2}$ with final state phases $\delta _{0}$ and $%
\delta _{2}$ as exhibted in Eqs. (\ref{e01}), (\ref{e02}) and (\ref{e03}).
Similar results follow by noting that the weak effective Lagrangian for $%
B\rightarrow \pi \pi $ decays contain both $\Delta I=1/2$ and $\Delta I=3/2$
parts. For $\Delta I=1/2$ $B\rightarrow \pi \pi $ transition the pions are
in $I=0$ or $I=1$ states where as for $\Delta I=3/2$ transition, they are in 
$I=1$ or $I=2$ states. On symmetrization, only two amplitudes $A_{0}$ and $%
A_{2}$ are left. The simple fact that there cannot be any other final state
phases except $\delta _{0}$ and $\delta _{2}$ allows us to express the
complex amplitudes $a_{0}$ and $a_{2}$%
\begin{equation}
a_{0}=f_{0}e^{i\theta _{0}}\text{, }a_{2}=f_{2}e^{i\theta _{2}}  \label{e04}
\end{equation}
where $\theta _{0}$ and $\theta _{2}$ are the weak phases. [Note this
parameterization would not have been possible for pions in $I=1$ state].

From the $CPT$ invariance, it follows that for the amplitudes $A_{+\,-}$ $%
A_{00}$ and $A_{+\,0}$, 
\begin{equation}
a_{0}\rightarrow a_{0}^{*}\text{, }a_{2}\rightarrow a_{2}^{*}  \label{e05}
\end{equation}
in Eqs. (\ref{e01}), (\ref{e02}) and (\ref{e03}). From these equations, we
obtain 
\begin{eqnarray}
\frac{R_{+\,\,-}}{R_{+\,0}} &=&\frac{\left| A\left( B^{0}\rightarrow \pi
^{+}\pi ^{-}\right) \right| ^{2}+\left| A\left( \bar{B}^{0}\rightarrow \pi
^{+}\pi ^{-}\right) \right| ^{2}}{2\left| A\left( B^{-}\rightarrow \pi
^{0}\pi ^{-}\right) \right| ^{2}}  \nonumber \\
&=&\frac{2}{9}\left[ 1+f^{2}+2f\cos \theta \cos \delta \right] \equiv \frac{2%
}{9}F  \label{e06} \\
\frac{R_{00}}{R_{-\,0}} &=&\frac{2}{9}\left[ 2+\frac{1}{2}f^{2}-2f\cos
\theta \cos \delta \right] \equiv \frac{2}{9}G  \label{e07} \\
\frac{R_{+\,\,-}+R_{00}}{R_{+0}} &=&\frac{2}{3}\left( 1+\frac{1}{2}%
f^{2}\right)  \label{e08} \\
C_{\pi ^{+}\pi ^{-}} &=&\frac{\left| A\left( B^{0}\rightarrow \pi ^{+}\pi
^{-}\right) \right| ^{2}-\left| A\left( \bar{B}^{0}\rightarrow \pi ^{+}\pi
^{-}\right) \right| ^{2}}{\left| A\left( B^{0}\rightarrow \pi ^{+}\pi
^{-}\right) \right| ^{2}+\left| A\left( \bar{B}^{0}\rightarrow \pi ^{+}\pi
^{-}\right) \right| ^{2}}  \nonumber \\
&=&-\frac{2f}{F}\sin \theta \sin \delta  \label{e09} \\
C_{\pi ^{0}\pi ^{0}} &=&\frac{2f}{G}\sin \theta \sin \delta  \label{e10}
\end{eqnarray}
where we have put 
\begin{equation}
f=-\sqrt{2}\frac{f_{0}}{f_{2}},\,\,\,\theta =\theta _{2}-\theta
_{0},\,\,\,\delta =\delta _{0}-\delta _{2}  \label{e11}
\end{equation}
From Eqs. (\ref{e06}), (\ref{e07}), (\ref{e09}) and (\ref{e10}), we get 
\begin{eqnarray}
\cos \theta \cos \delta &=&\frac{1}{6f}\left( F-2G+3\right) \equiv a
\label{e12} \\
\sin \theta \sin \delta &=&-\frac{F}{2f}C_{\pi ^{+}\pi ^{-}}\equiv b
\label{e13}
\end{eqnarray}
Before we proceed further, we note that with $f>0$, $\cos \theta \cos \delta 
$ must be positive, since $F>G$. Therefore, either $\theta $ and $\delta $
both lie in first quadrant or both lie in second quadrant. Thus we have 
\begin{equation}
0\leq \left| \theta -\delta \right| \leq 90^{0}  \label{e14}
\end{equation}
Without loss of generality, we can take $\theta $ and $\delta $ in the first
quadrant.

From Eqs. (\ref{e12}) and (\ref{e13}), we get 
\begin{eqnarray}
\cos \left( \theta +\delta \right) &=&a-b  \label{e15} \\
\cos \left( \theta -\delta \right) &=&a+b  \label{e16}
\end{eqnarray}
Thus $\theta $ and $\delta $ can be determined from the experimental values
of the parameters $a$ and $b$. However there is an ambiguity whether $\theta
<\delta $ or $\theta >\delta $. Here we take $\theta <\delta $. Also the
parameters $a$ and $b$ can be constrained as follows 
\begin{eqnarray}
-1 &\leq &a-b\leq 1  \nonumber \\
0 &\leq &a+b\leq 1  \label{e17}
\end{eqnarray}
Thus we get [noting from Eq. (\ref{e13}) that $b\geq 0$] 
\begin{equation}
0\leq b\leq 1-a\Rightarrow -\frac{2f}{F}\left( 1-a\right) \leq C_{\pi
^{+}\pi ^{-}}\leq 0  \label{e18}
\end{equation}
If $C_{\pi ^{+}\pi ^{-}}=0$, then it follows from Eqs. (\ref{e15}), and (\ref
{e16}) that either $\theta =0$ or $\delta =0$. This is what one would expect
for vanishing of direct $CP$-violation. There is no reason for $\delta =0$,
thus we choose the solution 
\begin{equation}
\theta =0\text{, }\delta =\cos ^{-1}a  \label{e19}
\end{equation}
On the other hand for the lower limit viz. $C_{\pi ^{+}\pi ^{-}}=-\left(
2f/F\right) \left( 1-a\right) $, we get 
\begin{eqnarray}
\theta -\delta &=&0\,\,\,\,\,or\,\,\,\,\,\,\,\theta =\delta  \nonumber \\
\cos 2\theta &=&2a-1  \label{e20}
\end{eqnarray}

Now using the experimental values for the branching ratios \cite
{PRL89-201802,PRD66-071102,Jawahery2003}: 
\begin{eqnarray*}
R_{+-} &=&\left( 4.6\pm 0.4\right) \times 10^{-6} \\
R_{00} &=&\left( 1.9\pm 0.6\right) \times 10^{-6} \\
R_{-0} &=&\left( 5.3\pm 0.8\right) \times 10^{-6}
\end{eqnarray*}
we get 
\begin{eqnarray}
f &=&1.3\pm 0.5,\,\,\,\,F=3.9\pm 0.7,\,\,\,\,\,G=1.6\pm 0.6  \nonumber \\
a &=&0.47\pm 0.23,\,\,\,\,\,\,\frac{f}{2F}=0.67\pm 0.26  \label{e21}
\end{eqnarray}
Hence from Eq. (\ref{e18}), we get the following bound on the direct
CP-asymmetry: 
\begin{equation}
-0.35\pm 0.22\leq C_{\pi ^{+}\pi ^{-}}\leq 0  \label{e22}
\end{equation}
Equations (\ref{e09}) and (\ref{e10}) give 
\begin{equation}
C_{\pi ^{0}\pi ^{0}}=\frac{F}{G}\left( -C_{\pi ^{+}\pi ^{-}}\right) =-\left(
2.4\pm 1\right) C_{\pi ^{+}\pi ^{-}}  \label{e23}
\end{equation}
The lower bound is remarkably close to the BaBar and Belle average \cite
{Jawahery2003}: $-0.38\pm 0.16$.

As noted above 
\begin{equation}
\theta =0,\,\,\,\delta =\cos ^{-1}a\Rightarrow 46^{0}\leq \delta \leq 76^{0}
\label{e24}
\end{equation}
On the other hand for the lower limit on $C_{\pi ^{+}\pi ^{-}}$, we have 
\begin{equation}
\theta =\delta ,\,\,\,\cos 2\theta =2a-1\Rightarrow 33^{0}\leq \theta \leq
61^{0}  \label{e25}
\end{equation}
To conclude this section, we note that direct $CP$-violation is constrained
between the parameter $-\left( 2f/F\right) \left( 1-a\right) $ [which can be
determined from the experimental values of the branching ratios $R_{+\,-}$, $%
R_{00}$ and $R_{-\,0}$]. Both the weak phase $\theta $ and the strong phase $%
\delta $ can be determined from the experimental values of $C_{\pi ^{+}\pi
^{-}}$ and the branching ratios. Equations (\ref{e23}), (\ref{e24}), (\ref
{e25}) and (\ref{e26}) are the main result of this section.

We now discuss, the mixing induced CP-asymmetry $S$. From Eqs. (\ref{e01}--%
\ref{e04}), we get 
\begin{eqnarray}
FS_{\pi ^{+}\pi ^{-}} &=&f^{2}\sin \left( 2\theta _{2}+2\phi _{M}-2\theta
\right) +\sin \left( 2\theta _{2}+2\phi _{M}\right) +2f\cos \delta \sin
\left( 2\theta _{2}+2\phi _{M}-\theta \right)  \label{e26} \\
GS_{\pi ^{0}\pi ^{0}} &=&2\sin \left( 2\theta _{2}+2\phi _{M}\right) +\frac{1%
}{2}f^{2}\sin \left( 2\theta _{2}+2\phi _{M}-2\theta \right) -2f\cos \delta
\sin \left( 2\theta _{2}+2\phi _{M}-\theta \right)  \label{e27}
\end{eqnarray}
From Eqs. (\ref{e26}) and (\ref{e27}), we can write an expression
independent of final phase $\delta $%
\begin{equation}
FS_{\pi ^{+}\pi ^{-}}+GS_{\pi ^{0}\pi ^{0}}=3\left[ \frac{1}{2}f^{2}\sin
\left( 2\theta _{2}+2\phi _{M}-2\theta \right) +\sin \left( 2\theta
_{2}+2\phi _{M}\right) \right]  \label{e28}
\end{equation}
We now discuss, two possible cases

\begin{enumerate}
\item  $C_{\pi ^{+}\pi ^{-}}=0$, $\theta =0$: 
\begin{eqnarray}
FS_{\pi ^{+}\pi ^{-}} &=&\left( 1+f^{2}+2f\cos \delta \right) \sin \left(
2\theta _{2}+2\phi _{M}\right)   \nonumber \\
S_{\pi ^{+}\pi ^{-}} &=&\sin \left( 2\theta _{2}+2\phi _{M}\right) 
\label{e29}
\end{eqnarray}

\item  $C_{\pi ^{+}\pi ^{-}}=-0.35\pm 0.22$, $\theta =\delta $, $33^{0}\leq
\theta \leq 61^{0}$: 
\begin{equation}
S_{\pi ^{+}\pi ^{-}}=\frac{1+f}{F}\left[ \sin \left( 2\theta _{2}+2\phi
_{M}\right) +f\sin \left( 2\theta _{2}+2\phi _{M}-2\theta \right) \right] 
\label{e30}
\end{equation}
With $\theta $ and $\delta $, determined from the experimental data as
outlined in the first section, it should be possible to determine $\left(
2\theta _{2}+2\phi _{M}\right) $ from the experimental values of $S_{\pi
^{+}\pi ^{-}}$, $S_{\pi ^{0}\pi ^{0}}$.
\end{enumerate}

Finally we note that $\phi _{M}=\beta $; and since $B^{\pm }\rightarrow \pi
^{\pm }\pi ^{0}$ decays are described by a single amplitude, we can identify 
$\theta _{2}$ with the angle $\gamma $ of CKM matrix. Hence we can rewrite
Eqs. (\ref{e29}) and (\ref{e30}) as 
\begin{eqnarray}
S_{\pi ^{+}\pi ^{-}} &=&-\sin 2\alpha  \label{e31} \\
S_{\pi ^{+}\pi ^{-}} &=&-\frac{1+f}{F}\left[ \sin 2\alpha +f\sin \left(
2\alpha +2\theta \right) \right]  \label{e32}
\end{eqnarray}

To Conclude: For $B\rightarrow \pi \pi $ decays, the pions must be in the
isospin symmetric states as required by Bose statistics. There are only two
isospin symmetric wave functions and hence there are only two amplitudes $%
A_{0}$ and $A_{2}$ with only two final state phases $\delta _{0}$ and $%
\delta _{2}$. Hence these decays can be analyzed in terms of one strong
phase $\delta =\delta _{0}-\delta _{2}$ and two weak phases $\theta $ and $%
\theta _{2}$. The direct $CP$-violation parameter $C_{\pi ^{+}\pi ^{-}}$ is
independent of $\theta _{2}$ and depends only on phases $\delta $ and $%
\theta $. These phases can be obtained from the branching ratios $R_{+\,-}$, 
$R_{00},$ $R_{-\,0}$ and $C_{\pi ^{+}\pi ^{-}}$. The mixing induced $CP$%
-violation parameters $S_{\pi ^{+}\pi ^{-}}$ in addition to $\theta $, $%
\delta $ and the mass mixing phase $\phi _{M}$ also contains the weak phase $%
\theta _{2}$. Having determined $\theta $ and $\delta $ from $C_{\pi ^{+}\pi
^{-}}$, the dependence of $S_{\pi ^{+}\pi ^{-}}$ reduces to two phases $%
\theta _{2}$ and $\phi _{M}$.

Acknowledgments: This work was supported in part by a grant from Pakistan
Council for Science and Technology.

\end{document}